\documentclass[prb,aps,amsmath,amssymb,showpacs,twocolumn]{revtex4-1}
\usepackage{graphicx,amsmath,amssymb,bm,xcolor,upgreek}
\usepackage[utf8]{inputenc}
\usepackage{nicefrac}
\usepackage{multirow, makecell, ragged2e}
\usepackage{diagbox}
\usepackage[titletoc,title]{appendix}
\usepackage{chngcntr}
\usepackage{float}

\newcommand{\etal}{\textit{et al}.}
\newcommand{\ie}{\textit{i}.\textit{e}.}
\newcommand{\eg}{\textit{e}.\textit{g}.}

\newcommand*{\rom}[1]{\expandafter\@slowromancap\romannumeral #1@}

\begin{document}
	
	\title{Fixed-phase diffusion Monte Carlo study of the activation gap and the skyrmion excitation of $\nu=1$ system in the presence of charged impurities}
	\author{Tongzhou Zhao}
	
	\affiliation{Department of Physics, 104 Davey Lab, Pennsylvania State University, University Park, Pennsylvania 16802, USA}

	\date{\today}
	
	\begin{abstract} 
The discrepancy between the theoretically calculated and experimentally measured activation gaps in quantum Hall effect has long been a puzzle. We revisit this issue in the context of the $\nu=1$ quantum Hall state, while also incorporating the skyrmion physics.We find that the finite width and the Landau level mixing (LLM) effects are not sufficient to explain the  observed activation gap. We further show that the presence of charged impurities located adjacent to the quantum well can cause a significant reduction in the activation gap, while also causing a suppression of the skyrmion size.
	\end{abstract}
	\maketitle
	
	\section{Introduction}
	Ever since the discovery of the fractional quantum Hall effect (FQHE)\cite{Stormer99,Tsui82}, many theoretical achievements have been made\cite{Laughlin81,Jain89} and a lot of experiment discoveries are successfully explained. However, a lot of questions remain. One of them is to quantitatively understand the excitation gap of FQHE states. The experimentally measured gaps\cite{Boebinger85,Willett88,Du93,Park99b,Scarola02,Pan20, Rosales21} are found to be significantly lower than the theoretically predicted ones\cite{Jain07,Morf86b,Haldane85a,Girvin85,Ortalano97,Park99}. We note here that even for the filling factor $\nu=1$, the theoretical value of the gap to charged excitations\cite{Fertig97,Cooper97,Melik-Alaverdian00,Melik-Alaverdian99} is much greater than the values observed in experiments\cite{Shukla00,Schmeller95}. That is the discrepancy we address in this work, with the belief that an understanding of it will help resolve the discrepancy for the fractional quantum Hall gaps.
	
	%At filling $\nu=1$, the ground state is fully polarized, as one might have expected from Hund's rule. 
	In the non-interacting picture, the ground state is fully polarized and the low energy excitation of the system in the small Zeeman energy limit consists of a quasi-particle and a quasi-hole pair, where a single spin-up electron is flipped and added to the lowest Landau level of the opposite spins. The energy of this excitation is modified substantially due to interaction, thus producing a gap that is much larger than the Zeeman splitting. There is another effect that lowers the gap, which has to do with the formation of the so-called skyrmion excitations. This appears most dramatically at zero Zeeman energy. Here, the ground state is still fully spin polarized due to exchange effects. However, exact diagonalization studies by Rezayi\cite{Rezayi87, Rezayi91a}  showed that even the removal of a single electron from the ground state (or adding one more electron into the system)  makes the system a spin singlet. This spin-singlet state was later identified by Sondhi \etal\cite{Sondhi93} to be the skyrmion state. This has lower energy than the state involving the flip of the spin of a single electron. For finite Zeeman energies, a skyrmion of a finite size is obtained\cite{Rezayi91a,Sondhi93,Xie96,Cooper97,Fertig97,Wojs02}.
	
	The activation gaps predicted by the skyrmion physics\cite{Fertig97, Cooper97} are significantly larger than experimental values measured in GaAs quantum wells\cite{Shukla00, Schmeller95}. Therefore, it requires further study to investigate what  factors  suppress the gap in realistic conditions.
	
	Many articles have tried to explain what lead to the reduction in the activation gap. Earlier Hartree-Fock calculations show that the finite width and the LLM can lower the skyrmion gaps\cite{Fertig97,Cooper97,Melik-Alaverdian00}. A 2D fixed-phase diffusion Monte Carlo (DMC) study by Melik-Alaverdian\etal\cite{Melik-Alaverdian99} shows that while either the finite width or the LLM reduces the gap, the LLM has a much weaker influence for finite width systems.  While all these studies show that the LLM and the finite width effect are responsible for the reduction of the activation gap, there is still a difference between the experimental gap and the theoretical gap including the correction of all these effects. A disorder-averaged Hartree-Fock calculation performed by Murthy shows that the effect of disorder may play a key role in reducing the excitation gap\cite{Murthy01}. Another study by Wan \etal  based on the Calculation of Chern number\cite{Wan05a} also shows the significance of disorder. These studies show that the activation gap is generally reduced in the presence of disorders, with or without the skyrmion physics. However, to the best of our knowledge, there is no work so far that includes the influence of the finite width, the LLM and disorder altogether. Therefore, a model that quantitatively captures all these factors in a realistic manner is needed.
	
	In this article, we report on our 3D fixed-phase DMC study of the skyrmion gap of $\nu=1$ system in the presence of charged impurities. Our study includes the LLM, the finite width effect and the influence of charged impurities simultaneously. Our calculation shows that without any impurities, the LLM and the finite width effect are not sufficient to reduce the activation gap to the value that has been observed in experiments. On the other hand, within our model, charged impurities that interact with the system by Coulomb interaction lead to an additional reduction in the activation gap. The calculated gap in the presence of charged impurities agrees with experiments reasonably well, given that all the characteristic parameters of impurities are in the range that is allowed by standard GaAs quantum well experiments\cite{Shukla00, Schmeller95}. We also find that the charged impurities suppress the size of skyrmions.
	
	In the following sections, we first review the background of the skyrmion physics at $\nu=1$. We next show our variational Monte Carlo (VMC) as well as DMC studies of the change of the gap due to the finite width and the LLM effect without any impurities. Finally, we show how the charged impurities influence the gap, and compare our results with experiments.
	\section{$\nu=1$ skyrmion}
	In this section, we briefly review the background of the skyrmion physics. Throughout this article, we specify the unit of length to be the magnetic length $l_B=\sqrt{\frac{\hbar c}{e B}}$ and the unit of energy to be the Coulomb energy $e^2/\varepsilon l_B$. We also define the LLM parameter $\kappa$ to be the ratio between the Coulomb energy and the cyclotron energy $\kappa=\frac{e^2/\varepsilon l_B}{\hbar \omega_\text{c}}$, where $\omega_\text{c}=\frac{e B}{m c}$. We work with the spherical geometry, where a magnetic monopole of the strength $Q$ is placed at the center of the sphere to generate the uniform magnetic field on the surface, which produces a magnetic flux of $2Q hc/e$. Before we introduce the skyrmion wave functions, it is necessary to first give the ground state wave function for the quantum Hall state at $\nu=1$ and the quasi-particle as well as the quasi-hole wave functions. The ground state wave function at $\nu=1$ is obtained by filling up all the lowest Landau level orbitals. This requires the number of electrons $N$ and the magnetic strength $Q$ to satisfy $2Q+1=N$. The ground state wave function (unnormalized) reads
	\begin{equation}\label{wfn_gs}
	\begin{aligned}
	\Psi_{\text{gs};N}=\prod_{1\leq j<k\leq N}(u_j v_k -u_k v_j)
	\end{aligned}
	\end{equation}
	where $u_i = \cos\left(\theta_i/2\right)e^{i \phi_i/2}$ and $v_i=\sin\left(\theta_i/2\right)e^{-i \phi_i/2}$ are spinor coordinates\cite{Haldane83,Wu76} of the $i$'th particle, where $\theta_i$ and $\phi_i$ are the usual spherical angles. The quasi-hole wave function is obtained by removing an electron from the ground state at the north pole of the sphere by setting $u_N=0$ and $v_N=1$:
	\begin{equation}\label{wfn_hole}
	\begin{aligned}
	\Psi_{\text{h};N-1}=\prod_{1\leq j<k<N}(u_j v_k -u_k v_j) \prod_{1\leq j<N}u_j.
	\end{aligned}
	\end{equation}
	We note that because for the ground state and the quasi-hole wave functions, only the spin-up orbitals are occupied, we do not write down the spins explicitly. We will write down the spin part of the wave functions explicitly when we give wave functions for the quasi-particle state and the skyrmion state to avoid any confusion since those two states involve both spins.
	
	To calculate the excitation gap, we also need the wave function for the quasi-particle state. The wave function is constructed by adding an electron to the ground state at the south pole of the sphere:
	\begin{equation}\label{wfn_particle}
	\begin{aligned}
	\Psi_{\text{p}, N+1}=\mathcal{A}\left[\Psi_{1, N}Y_{QQ(-Q)}\right]\uparrow_1...\uparrow_{N} \downarrow_{N+1}]
	\end{aligned}
	\end{equation}
	where $\mathcal{A}$ is the antisymmetrization operator, $\uparrow$ and $\downarrow$ denote up- and down-spins, and $Y_{QQ(-Q)}$ is the magnetic harmonic in the LLL with the maximum quantum number for $L_z$\cite{Jain07}:
	\begin{equation}
	Y_{QQ(-Q)}=[\frac{2Q+1}{4\pi}]^{1/2}v^{2Q}
	\end{equation}
	
	We use the skyrmion wave function on the sphere proposed by MacDonald, Fertig and Brey\cite{MacDonald96}. The skyrmion wave function for $N-1$ particles with $K+1$ spins flipped relative to the ground state is given by:
	\begin{equation}\label{wfn_skyrmion}
	\begin{aligned}
	\Psi_{\text{sk};N-1}^K &=\Psi_{\text{gs}, N-1} \sum_{\{i_1, ...,i_K\}}\left[ v_{i_1}...v_{i_K}u_{j_1}...u_{j_{N-1-K}} \cdot\right.\\
	&\cdot\left.\downarrow_{i_1}...\downarrow_{i_K}\uparrow_{j_1}...\uparrow_{j_{N-1-K}}\right]
	\end{aligned}
	\end{equation}
	where the sum is over all distinct particle indices and $j$s denote the particles other than $i_1, i_2, ..., i_K$. When $K=0$, the above wave function becomes the quasi-hole wave function, \ie, $\Psi_{\text{h};N-1}=\Psi_{\text{sk};N-1}^{K=0}$, as expected.
	
	In principle, one can also obtain the wave function for the anti-skyrmion by applying the particle-hole(PH) conjugation on the skyrmion wave function in the Fock space. However, we do not include this state in our study because of the reasons that we are going to discuss shortly.
	
	The gap for flipping a single particle is defined as:
	\begin{equation}\label{gap_ph}
	\begin{aligned}
	\Delta_{\text{ph}}=E_\text{p}+E_\text{h}-2E_\text{gs}+E_\text{Z}.
	\end{aligned}
	\end{equation}
	Here $E_\text{p}$ is the total energy of the quasi-particle state with $N+1$ particles, $E_\text{h}$ is the total energy with $N-1$ particles and $E_\text{gs}$ is the ground state energy of $N$ particles. The gap of the skyrmion-anti-skyrmion pair of $S=2K+1$ total spins flipped is defined as:
	\begin{equation}\label{gap_Sk_aSk}
	\begin{aligned}
	\Delta_\text{sk-ask}(K)=E_\text{sk}(K)+E_\text{ask}(K)-2E_\text{gs}+(2K+1) g.
	\end{aligned}
	\end{equation}	
	where $g$ is the Land\'{e} $g$ factor and the relationship that $E_\text{Z}=S g$ has been used.
	When the LLM is not considered, the PH symmetry is preserved. In this case the energies of generating a pair of quasi-hole-skyrmion and a pair of quasi-particle-anti-skyrmion are the same. Therefore one can rewrite Eq.~\ref{gap_Sk_aSk} as:
	\begin{equation}\label{gap_Sk_aSk_delta}
	\Delta_\text{sk-ask}(K) = \Delta_\text{ph}+2\delta(K).
	\end{equation}
	where $\delta(K) = E_\text{sk}-E_\text{h} = E_\text{ask}-E_\text{p}.$ On the other hand, when the PH symmetry is broken by the LLM, Eq.~\ref{gap_Sk_aSk_delta} does not hold anymore and one needs to explicitly calculate the gap of generating a pair of skyrmion and anti-skyrmion to obtain the proper excitation gap. However, the anti-skyrmion state is very difficult to handle in the DMC algorithm, which prevents us from performing an explicit calculation for anti-skyrmion state. Nonetheless, we will continue to use Eq.~\ref{gap_Sk_aSk_delta} even when LLM is present. This is because the LLM typically only makes a small modification on the energy and only breaks the PH symmetry slightly at experimental parameters\cite{Melik-Alaverdian97,Sreejith17,Zhao18,Zhao20}. Compared to the big energy discrepancy that we are concerned with throughout this article, Eq.~\ref{gap_Sk_aSk_delta} is still a very accurate estimation of the activation gap.

	The skyrmion size is determined by minimizing $\Delta_\text{sk-ask}(K)$ over all values of $K$. For the $N$-particle system, the maximum value of $K$ is $N/2$, which corresponds to the spin-singlet state. One needs to calculate the energies of all the skyrmion states and pick out the one that has the lowest $\Delta_\text{sk-ask}$.
	\section{Variational Monte Carlo study}
	In this section we introduce our VMC study of the problem. There are two different calculations that are performed. First, we neglect the finite width of the quantum well. Next, we perform a calculation that includes the finite width effect of the quantum well by taking the effective Coulomb interaction between particles as:
	\begin{equation}
	\begin{aligned}
	V_{\text{eff}}(r)=\int dz\frac{\rho(z)^2}{\sqrt{r^2+z^2}},
	\end{aligned}
	\end{equation}
	where $\rho(z)$ is the transverse distribution, which is evaluated in local density approximation at zero magnetic field\cite{Park99b,Martin20}. In our study, the width of the quantum well is sufficiently narrow that only the lowest subband is involved. (We choose the parameters similar to those in Ref.~\cite{Shukla00, Schmeller95} where the carrier densities are around $10^{11} \text{cm}^{-2}$ and the quantum well widths are around $20 \text{nm}$, which corresponds to a width of $1- 2 l_B$.) 
	
\begin{figure}[H]
		\includegraphics[width=\columnwidth]{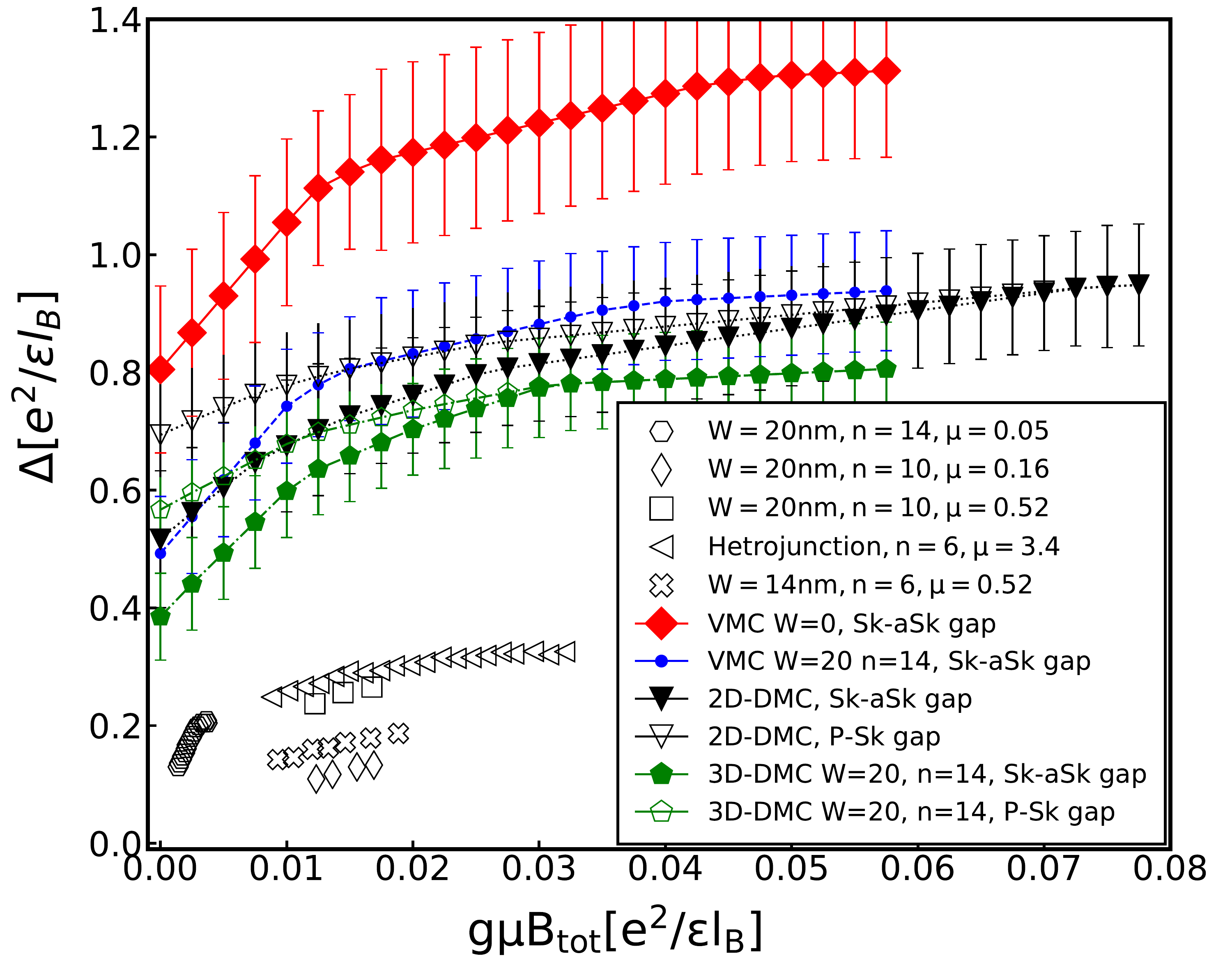}
		\includegraphics[width=\columnwidth]{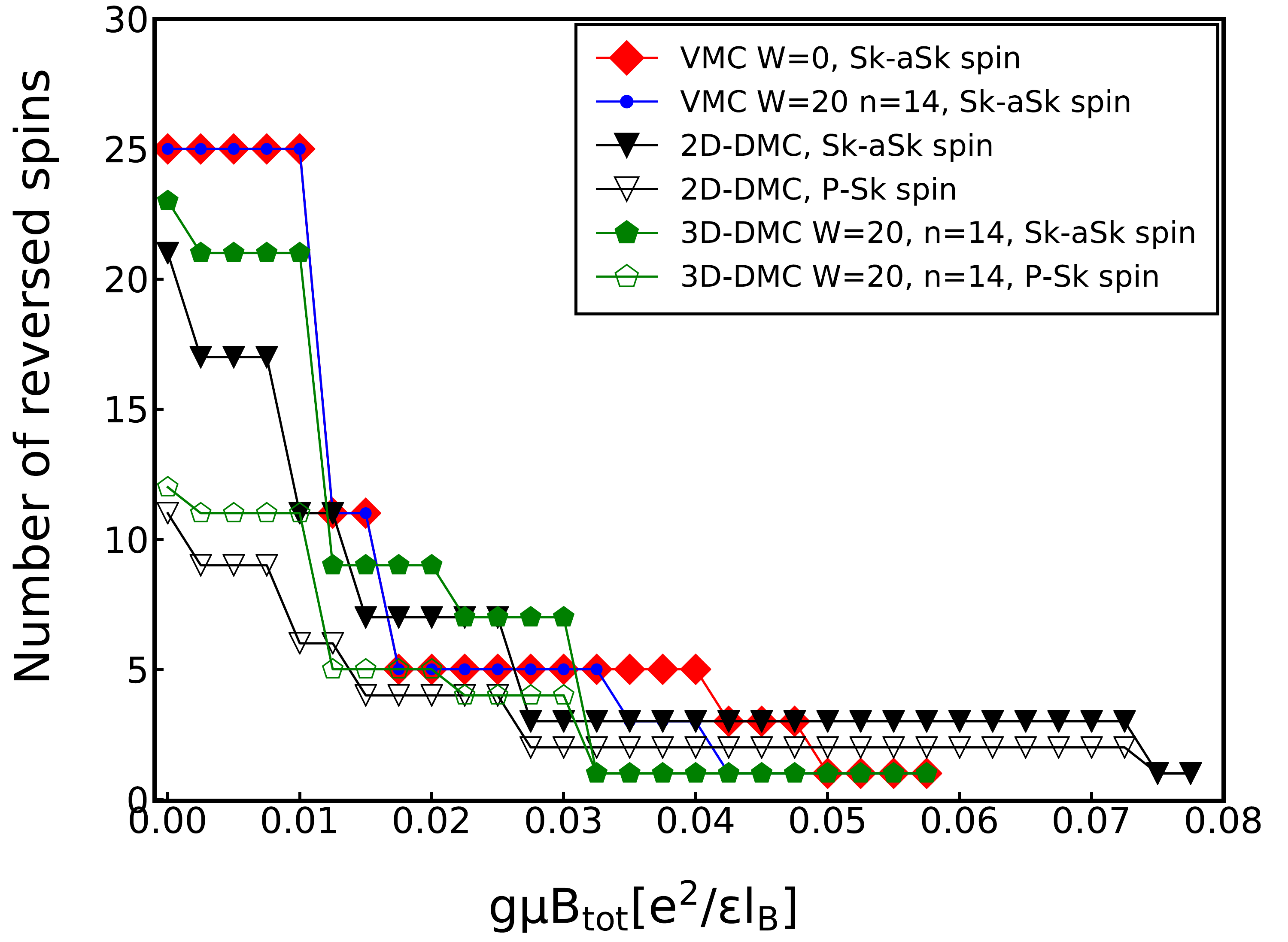}
		\caption{The gap and the total number of reversed spins calculated by VMC and DMC methods for skyrmion states at $\nu=1$ . Top: experimental and theoretical gaps at different parameters. Bottom: the total number of reversed spins calculated by VMC and DMC methods. In the legend, $W$ stands for the quantum well width in $\text{nm}$ and $n$ stands for the carrier density in $10^{10}\text{cm}^{-2}$. Experimental data are plotted with scattered marks with the mobilities labeled by $\mu$ in units of $10^6 \text{cm}^2/\text{V s}$.}
		\label{fig_energy_all}
	\end{figure}

	The results are shown in Fig.~\ref{fig_energy_all}. We find that the finite width effect can reduce the energy gap by about $20\%$ to $30\%$. The gap after the inclusion of finite width is still significantly greater than experimental values. We note that our results agree with earlier Hartree-Fock and Monte Carlo calculations\cite{Fertig97,Cooper97,Melik-Alaverdian00}. We also show the total number of reversed spins $S$ from our calculation. The smallest system has 24 particles so the largest $K$ considered in our calculation is $12$. This sets an upper limit of $S$ in our calculation to be $25$.
	\section{Diffusion Monte Carlo study}
	Because the VMC results suggest that the finite width itself is not sufficient to explain the reduction in the activation gap, we proceed to explore how LLM influences the gap. This is done by the fixed-phase DMC calculation. The fixed-phase DMC is a type of the DMC method and it is specialized to solve the many-body Schr\"odinger equation where the wave functions for the system cannot be written as real functions (\eg, FQHE systems where time reversal symmetry is broken). Essentially, this method approximates the phase of the ground state wave function by the phase of a well-defined trial wave function and uses the standard DMC technique to solve for the amplitude of the wave functions. It has been proven to be effective in solving FQHE problems\cite{Guclu05a,Ortiz93,Melik-Alaverdian95,Bolton96,Melik-Alaverdian97,Melik-Alaverdian99,Zhao18,Zhang16,Hossain20} and we also refer readers to Ref.~[\onlinecite{Reynolds82, Foulkes01}] for general introduction of the standard DMC. Here we only note that the fixed-phase DMC automatically include the effect of LLM.
	
	In Fig.~\ref{fig_energy_all} we also show our DMC results. First we show the 2D-DMC where the system is confined within the plane. Next we include the finite width of the quantum well by allowing the system to evolve in the real 3D space to minimize its energy. We find that LLM itself can reduce the gap by about $20\%\sim 30\%$, which is comparable to the reduction caused by finite width effect alone. When both LLM and finite width are considered, the gap is only further decreased by very little. The fact that the LLM has a much weaker effect for finite width systems agrees with the conclusion in Ref.~[\onlinecite{Melik-Alaverdian99}]. This is not very surprising, as the finite width effect softens the repulsion between particles, less mixing with higher Landau levels is needed.
	\section{The role of charged impurity}
\begin{figure}[H]
		\includegraphics[width=0.48\columnwidth]{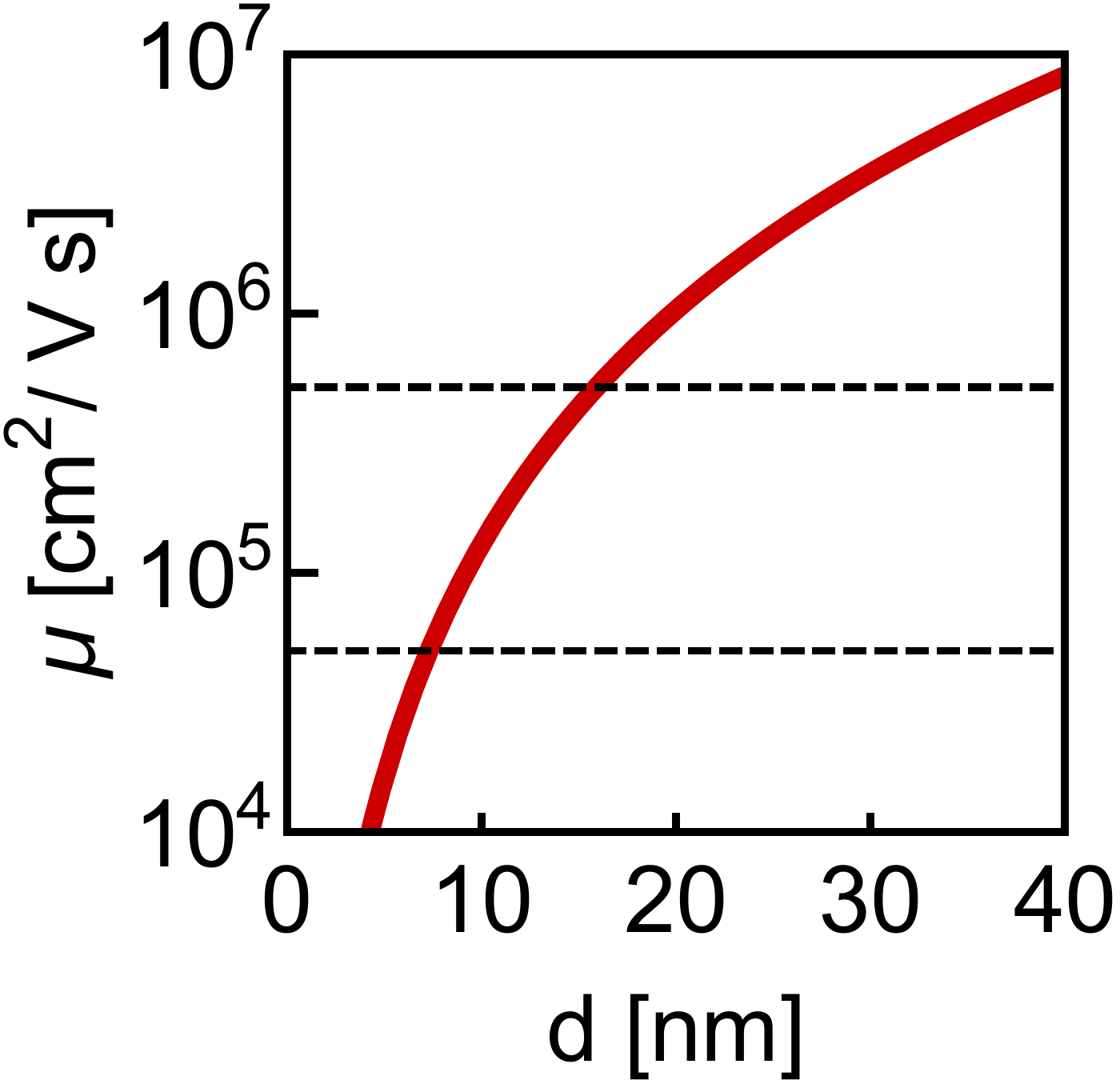}
		\includegraphics[width=0.48\columnwidth]{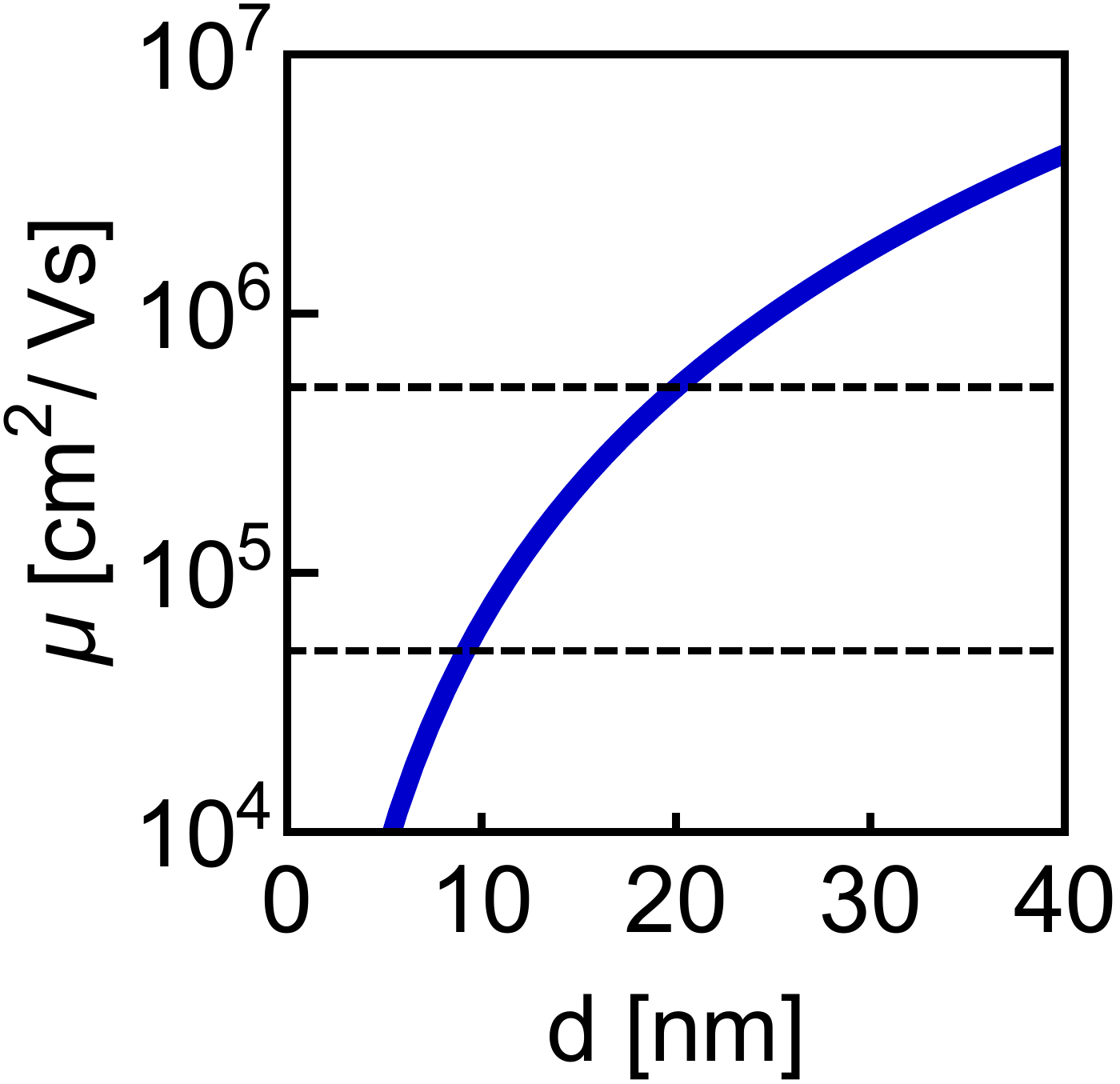}
		\caption{The mobility due to Coulomb scattering as a function of the impurity separation $d$, calculated with the Born approximation. Left: $Z=0.5$ and $n_\text{imp}=2.8\times 10^{11} \text{cm}^{-2}$. Right: $Z=1$ and $n_\text{imp}=1.4\times10^{11} \text{cm}^{-2}$. The dashed lines show the mobility values observed in Ref.~[\onlinecite{Shukla00}] ($\mu=5\times10^4 \text{cm}^2/\text{ V s},  n=1.4\times10^{11}\text{cm}^2/\text{ V s}$) and Ref.~[\onlinecite{Schmeller95}]($\mu=5\times10^5 \text{cm}^2/\text{V s}, n=10^{11}\text{cm}^{-2}$).}
		\label{mobility_fig}
	\end{figure}

	Based on our results from VMC and DMC, we find that the LLM and finite width are not sufficient to explain the discrepancy between theory and experiments. Another important factor that has not been included in our study is the influence of disorders. It is believed that at low temperatures, the scattering process is dominated by charged impurities via the Coulomb interaction\cite{Ando82}. This is partially justified by the fact that the theoretically calculated mobility due to the Coulomb scattering agrees with experiments\cite{Ando82a,Harrang85,Zheng96}. For our purpose, we consider a simple model: a negatively charged impurity $q$ is placed above the north pole of the sphere and a charged $-q$ impurity is placed above the south pole. We place the impurities at these locations because in general quasi-holes are attracted to negative charges and quasi-particles are attracted to positive charges. Since we add the quasi-hole at the north pole and the quasi-particle at the south pole, such a configuration should be energetically favored. We also note that because only one skyrmion-anti-skyrmion pair is inside our system, there is only one pair of charged impurities. This configuration is only suitable to study sparsely distributed impurities since each skyrmion is only affected by one impurity. We next specify the separation between the impurities and the quantum well to be some values of the order of a few hundred angstroms. The separation is chosen as such because experimentally the Coulomb impurities are typically introduced in doping region that is separated from the quantum well by a few hundreds of angstroms. We will see shortly that this is also supported by a quantitative calculation of the mobility through estimating the relaxation time ($\mu=\frac{e \tau}{m}$) due to the Coulomb scattering. In general, one needs a sophisticated model to properly include the effects from finite width of the quantum well, the distribution of impurities, and the screening effect on the permittivity\cite{Ando82,Davies98}. However, because we do not know all these details, we can only look for a semi-quantitative description of the disorder at best. The mobility of the 2D electron gas in the Born-approximation at zero temperature reads\cite{Davies98}:
	\begin{equation}\label{mobility}
		\mu=\frac{8 e (k_F d)^3}{Z^2 \pi \hbar n_\text{imp}},
	\end{equation}
	where $d$ is the separation between the impurity and the 2D system, $k_F$ is the Fermi wave vector, $Z e$ is the impurity charge and $n_\text{imp}$ is the 2D density of impurities. In our study, we make the assumption that the total charge of impurities is the same as the total charge of electrons, \ie, $Z n_\text{imp}=n$. In Fig.~\ref{mobility_fig}, we show that for $Z=0.5$ and $Z=1$, this model gives a mobility that agrees with the value measured in experiments\cite{Shukla00, Schmeller95} when $d$ is around $10 \text{nm}$ to $30 \text{nm}$.
	
\begin{figure*}
	\includegraphics[width=0.48\linewidth]{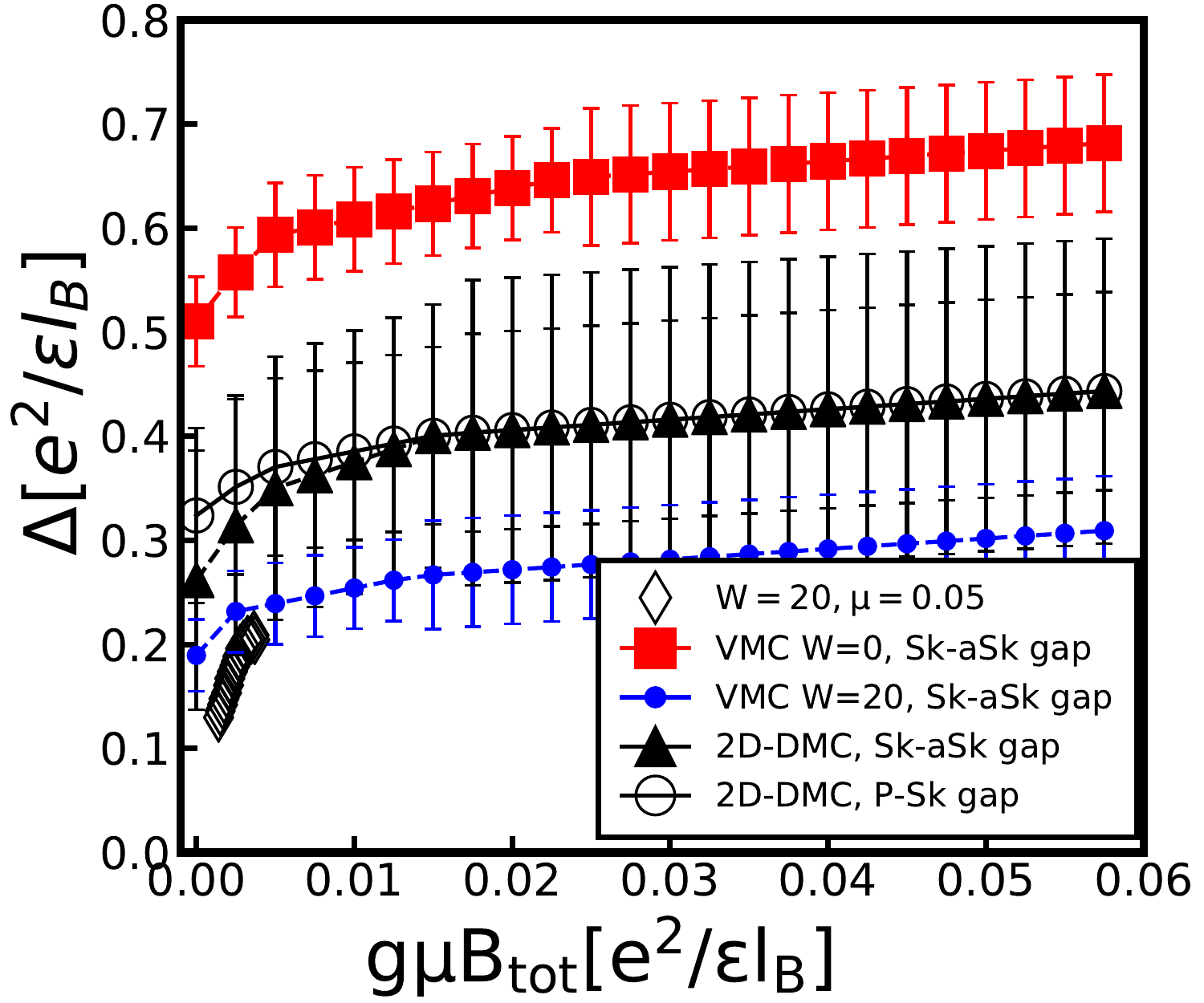}
	\includegraphics[width=0.48\linewidth]{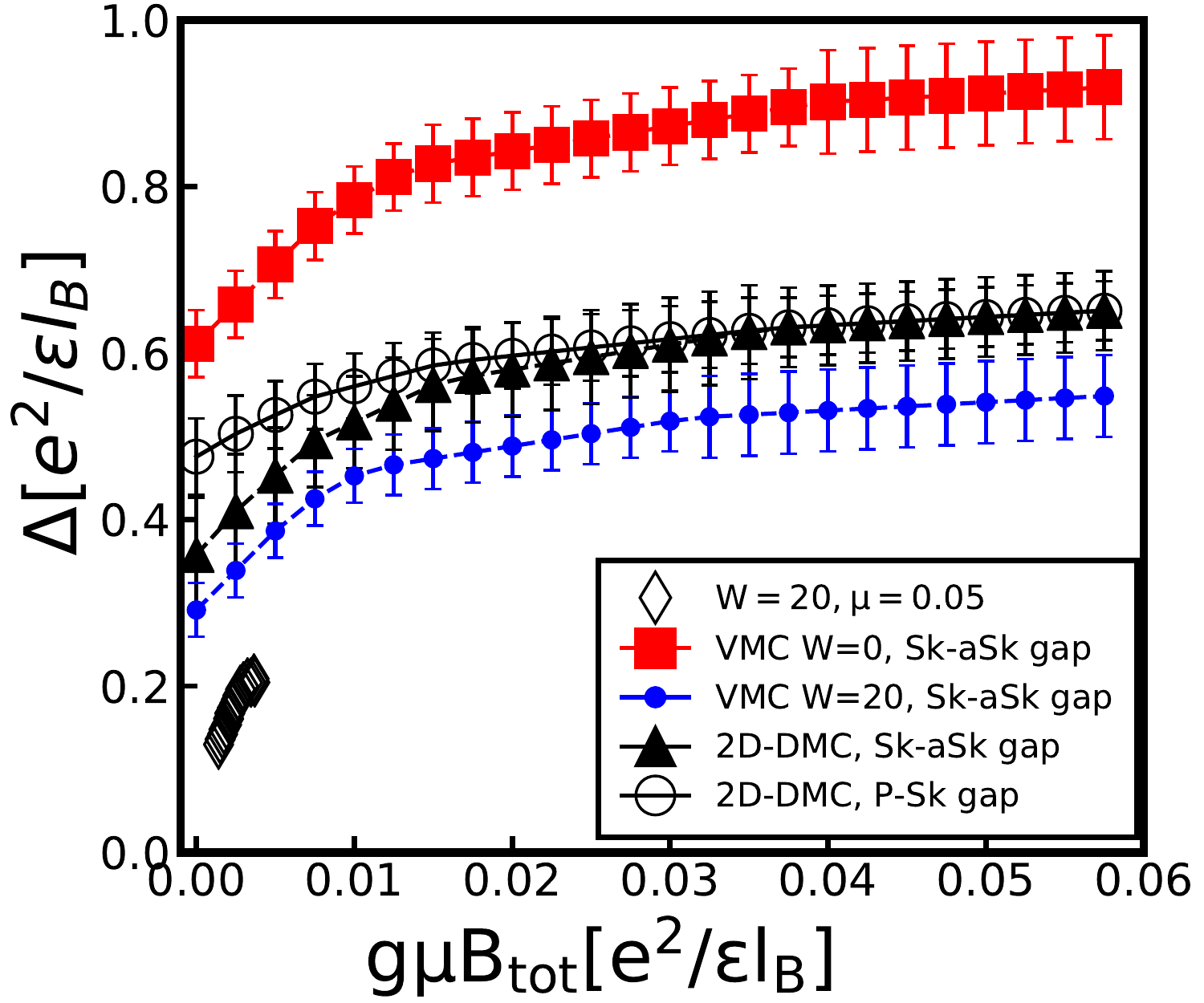}
	\includegraphics[width=0.48\linewidth]{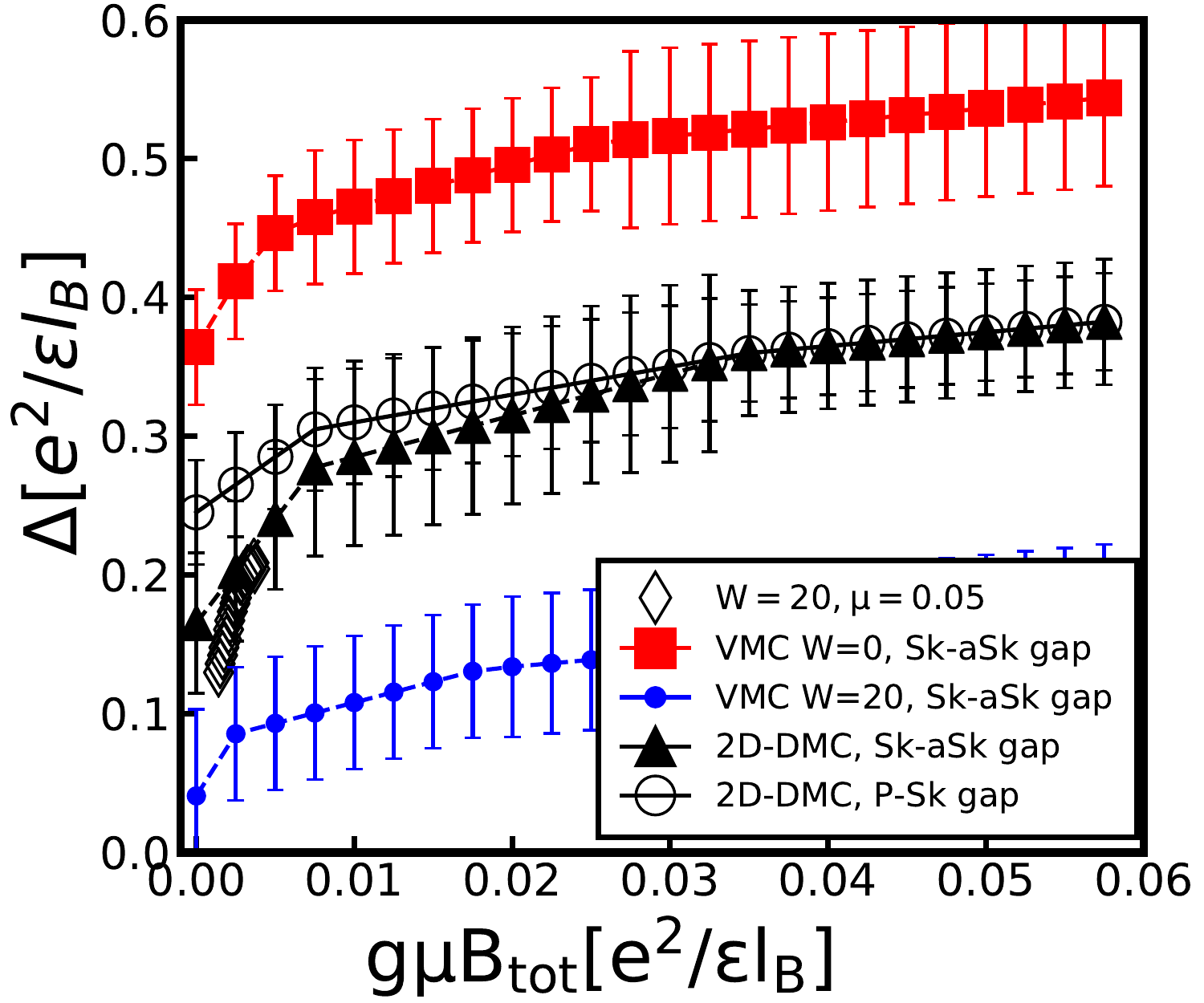}
	\includegraphics[width=0.48\linewidth]{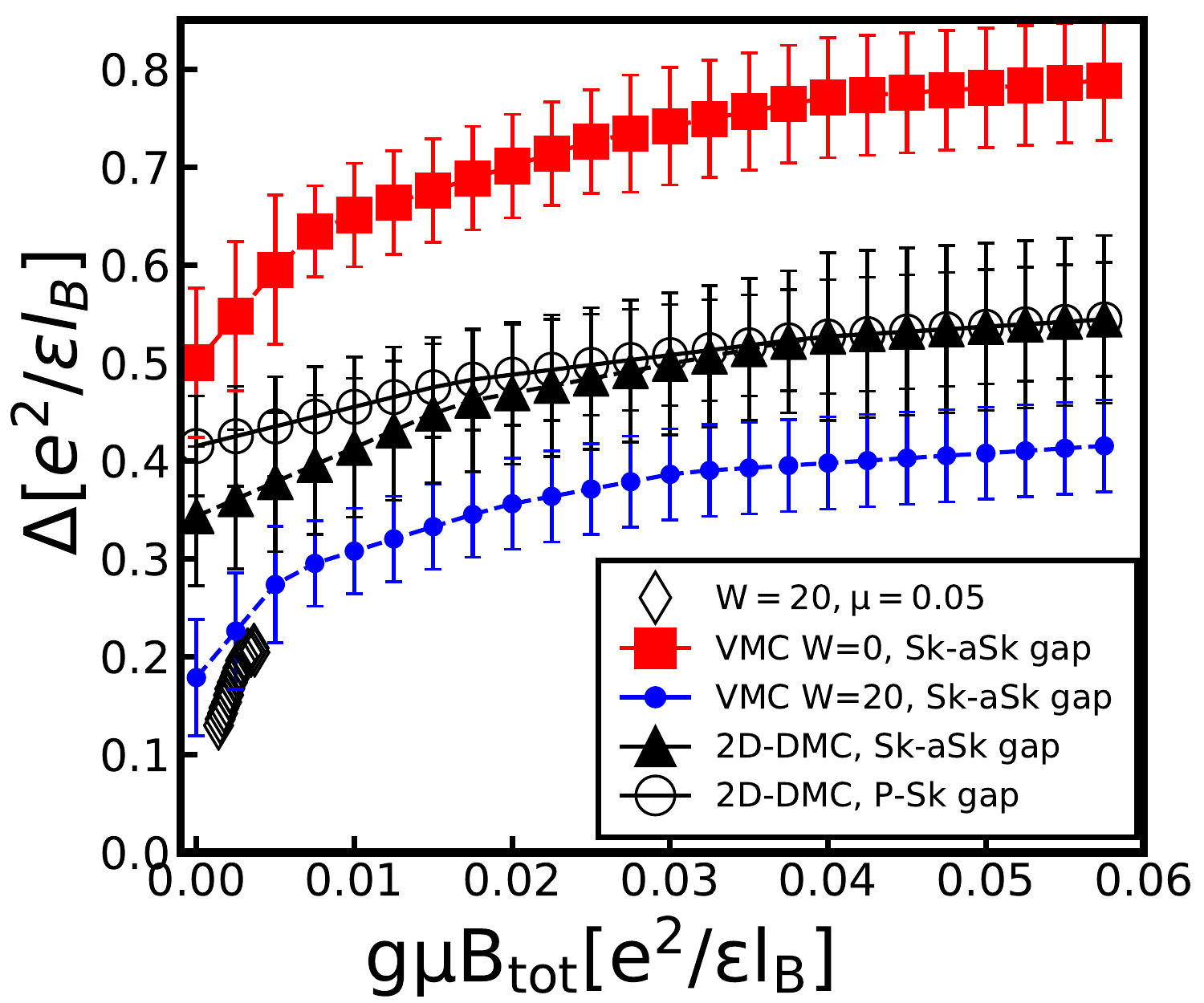}	
	\caption{The activation gap calculated by VMC and 2D-DMC methods for skyrmion at $\nu=1$ with charged impurity $q$ separated from the quantum well by the distance $d$. Top-left: $q=0.5 e$ and $d=10 \text{nm}$; Top-left: $q=0.5 e$ and $d=20 \text{nm}$; Bottom-left: $q=e$ and $d=20 \text{nm}$; Bottom-right: $q=e$ and $d=30 \text{nm}$. In the legend, $W$ stands for the quantum well width in $\text{nm}$ and $\mu$ is the mobility in $10^6 \text{cm}^{2}/\text{V s}$. The LLM of 2D-DMC is $\kappa=1$, which corresponds to $n=1.4\times 10^{11}\text{cm}^{-2}$. Experimental data are drawn with scattered points for comparison.}
	\label{fig_energy_impurity}
	\end{figure*}

\begin{figure*}
		\includegraphics[width=0.48\linewidth]{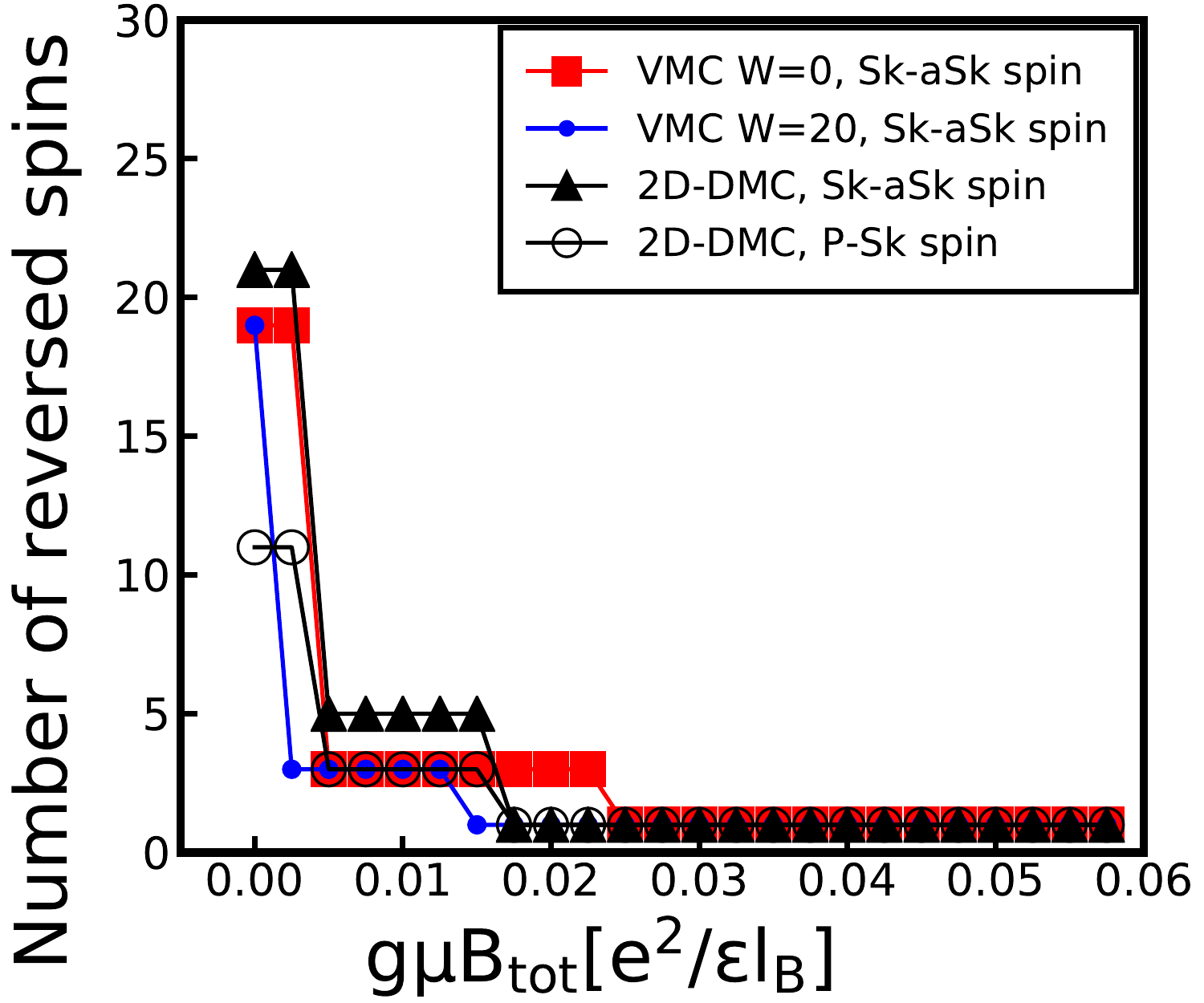}
		\includegraphics[width=0.48\linewidth]{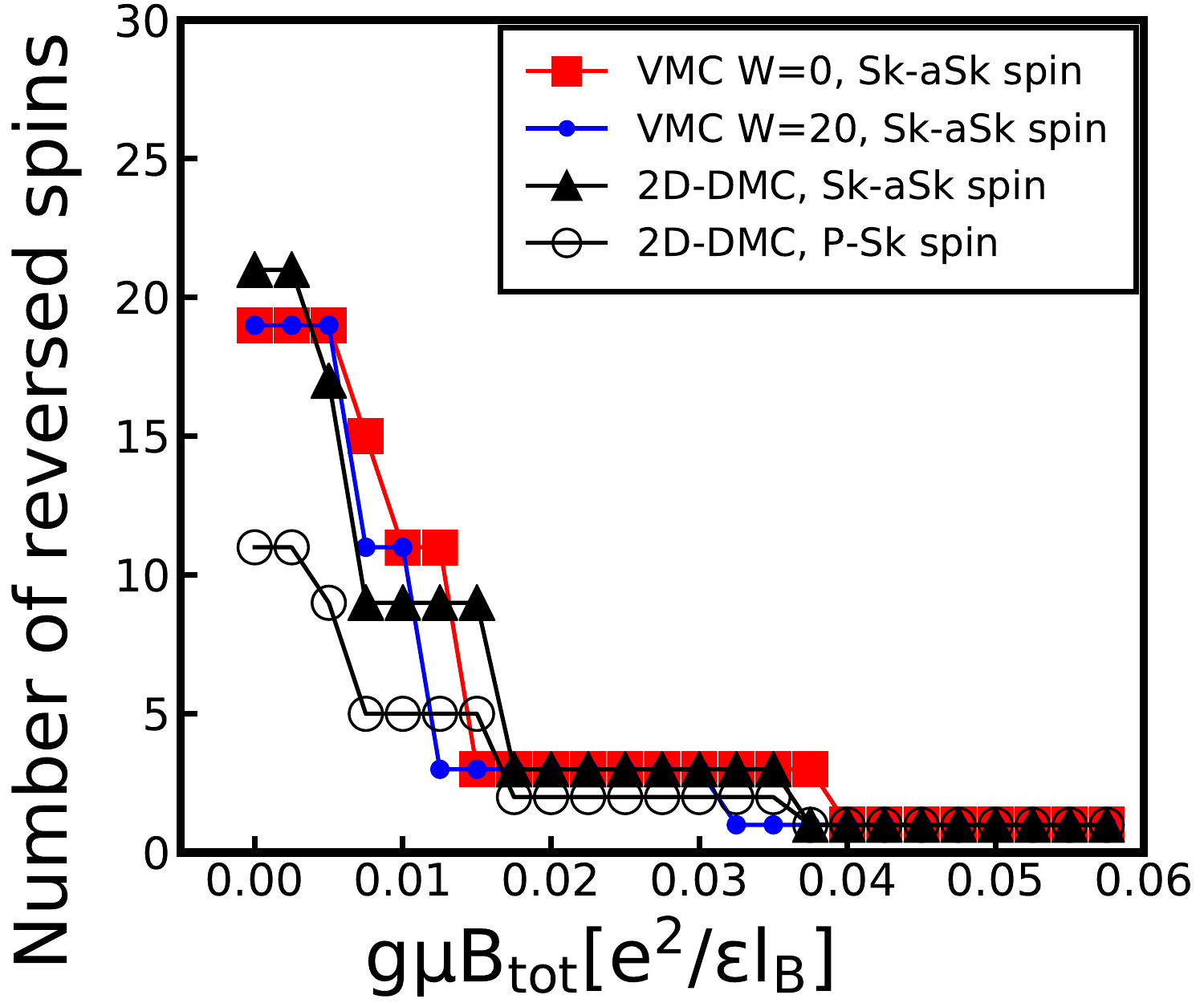}
		\includegraphics[width=0.48\linewidth]{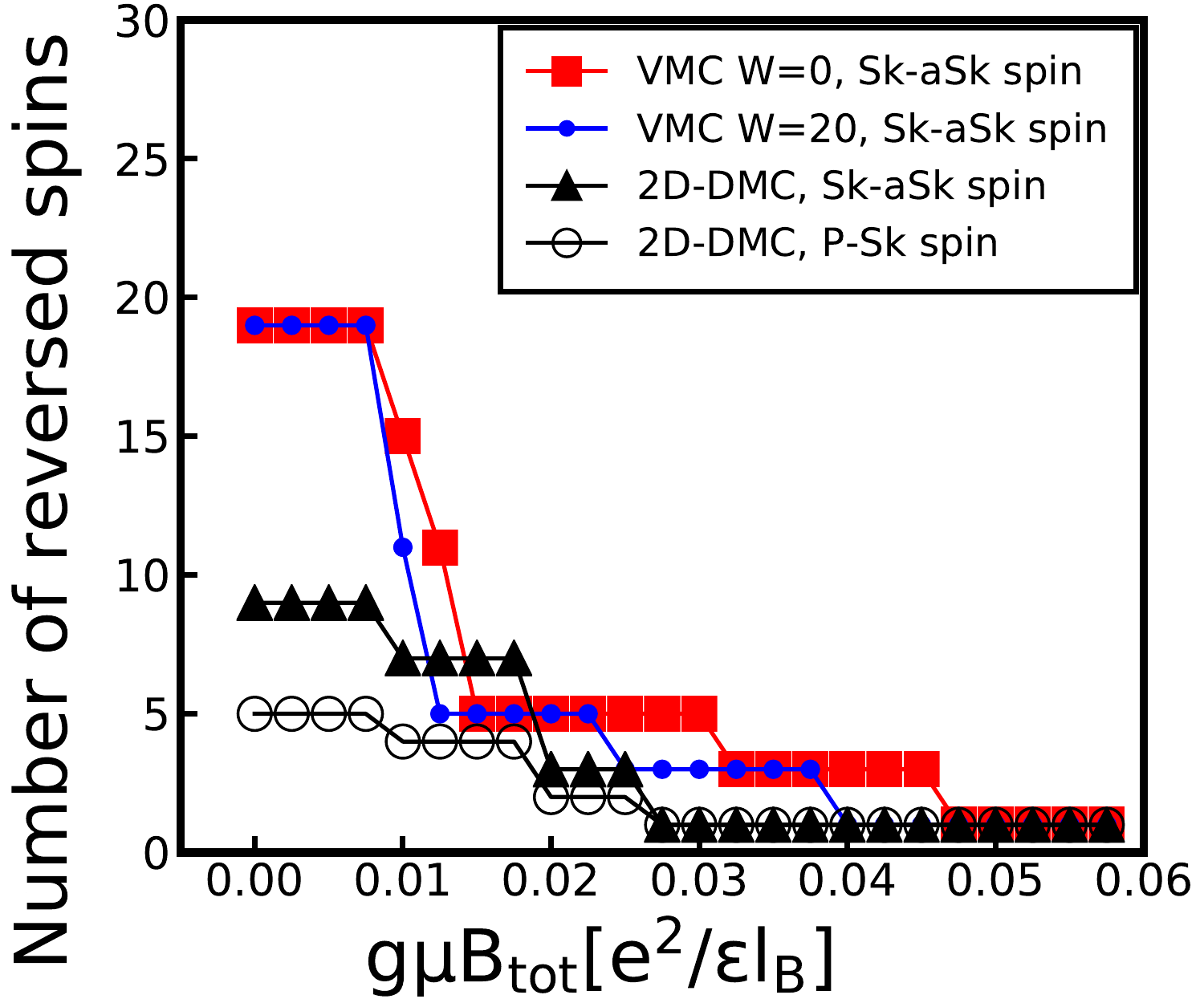}
		\includegraphics[width=0.48\linewidth]{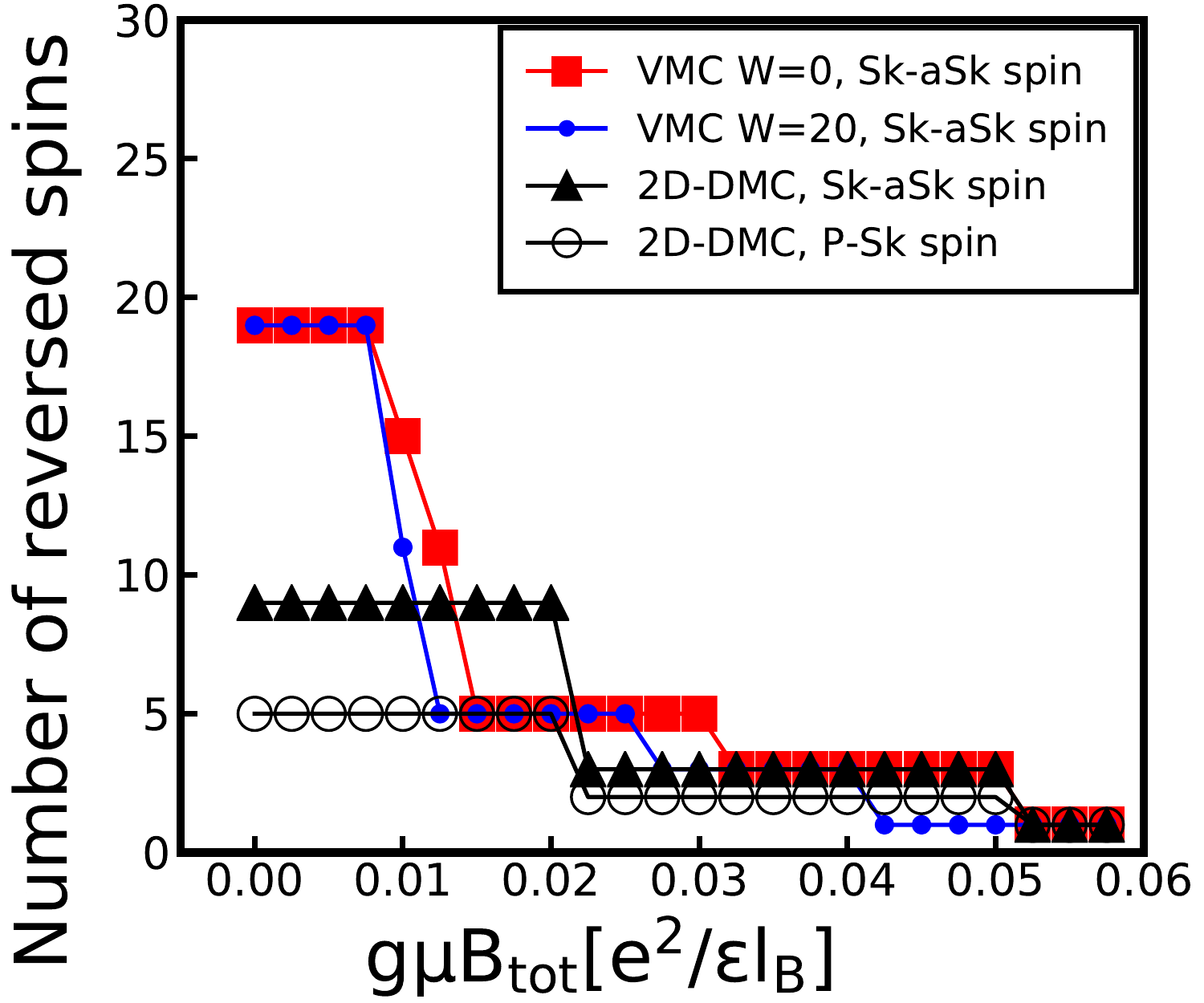}	
		\caption{The number of total reversed spins of skyrmions at different impuritiy separations $d$ calculated by VMC and 2D-DMC in the presence of charged impurties in the quantum well at $n=1.4\times 10^{11} \text{cm}^{-2}$. Top-left: $d=10 \text{nm}$; Top-rigth: $d=20 \text{nm}$; Bottom-left: $d=30 \text{nm}$; Bottom-right: $d=40 \text{nm}$.}
		\label{fig_spin_impurity}
	\end{figure*}

	The results from VMC and 2D-DMC calculation are shown in Fig.~\ref{fig_energy_impurity}. In general, we find that the inclusion of charged impurities leads to a much better agreement between the theoretical and the experimental values of the activation gap. We find that for both impurity charges ($q=e$ and $q=0.5e$), the skyrmion physics arises at small Zeeman energy and the gaps are comparable to experiments. Our calculation also gives the values of $S$ at different $g$s. Our calculation suggests that the skyrmion physics can survive a wide range of charged impurity strength and distance. Particularly, a system with a large impurity charge and a large impurity distance can coincide with a system with a small impurity charge and a small impurity distance in their gaps. This gives a possible explanation for why the samples from Ref.~[\onlinecite{Schmeller95}] and Ref.~[\onlinecite{Shukla00}]  have very similar excitation gaps while their mobilities differ by about ten times.
	Lastly we show that there is a suppression effect on the skyrmion size due to the presence of impurities. As one can see in Fig.~\ref{fig_spin_impurity}, when we fix the impurity charge to be $0.5e$, the occurring of the large size skyrmions ($S>2$) is delayed as the charge distance decreases. This can be understood by the fact that the skrymion with a large $K$ value has its charge distribution more extended, so the point-like impurity has a stronger attraction with small skyrmions. The calculation suggests that in order to obtain large-size skyrmions experimentally, the charged impurities should be separated from the quantum well distantly.
\section{Conclusion}
In this paper, we have revisited the question of the discrepancy between the theoretically calculated and experimentally measured activation gaps in the $\nu=1$ quantum Hall state. We have found that the puzzle cannot be resolved by only considering the finite width effect and the LLM effect. We have proposed a simple model to include the influence of sparsely distributed charged impurities and our conclusion is that the Coulomb impurities can greatly reduce the activation gap and they can also suppress the size of skyrmions. While our model can explain the experimental observations, it may lack realistic details, and thus our model is at its best a semi-quantitative account of the puzzle. More experiments are required to further elucidate this issue. 

\textit{Acknowledgement}: I am grateful to J. K. Jain for our discussion on the results and his advice on the manuscript. I would also like to thank Mansour Shayegan and Ganpathy Murthy for many insightful discussions. The work was made possible by financial support from the U.S. Department of Energy
under Award No. DE-SC0005042. The numerical calculations were performed using Advanced CyberInfrastructure computational resources provided by The Institute for CyberScience at The Pennsylvania State University. 
	%\bibliography{../../Latex-Revtex-Master/biblio_fqhe}
\end{document}